\documentclass[10pt,twocolumn,english,aps,prd,nofootinbib,showkeys,preprintnumbers,showpacs]{revtex4}
\usepackage[utf8]{inputenc}
\usepackage{graphicx}
\usepackage{amsmath}
\usepackage{amssymb}
\usepackage{amsfonts}
\usepackage{lscape}
\usepackage{hyperref}
\usepackage{url}
\usepackage{epsfig}
\usepackage{color}
\usepackage{bm}
\usepackage{tabularx}
\newcommand{\be}{\begin{equation}}
\newcommand{\ee}{\end{equation}}
\newcommand{\ba}{\begin{eqnarray}}
\newcommand{\ea}{\end{eqnarray}}
\newcommand{\br}{\begin{array}}
\newcommand{\er}{\end{array}}

\newcommand{\rr}{\mathrm}

\newcommand{\dd}{\partial}

\newcommand{\OM}{\Omega_{\rm M}}

\newcommand{\Msun}{M_\odot}

\newcommand{\rpbh}{\rho_{\rm PBH}}

\newcommand{\fpbh}{f_{\rm PBH}}

\newcommand{\e}[1]{_{\rm #1}}

\makeatother

\begin{document}

\title{LIGO Lo(g)Normal MACHO\footnote{It sounds even funnier in Spanish.}: Primordial Black Holes survive SN lensing constraints}

\author{Juan Garc\'ia-Bellido$^{(a,b)}$}\email{juan.garciabellido@cern.ch}
\author{S\'ebastien Clesse$^{(c,d)}$}\email{sebastien.clesse@unamur.be}
\author{Pierre Fleury$^{(e)}$}\email{pierre.fleury@unige.ch}
\affiliation{$^{(a)}$Instituto de F\'isica Te\'orica UAM-CSIC, Universidad Auton\'oma de Madrid,
Cantoblanco, 28049 Madrid, Spain\\
$^{(b)}$CERN, Theoretical Physics Department, 1211 Geneva, Switzerland\\
$^{(c)}$ Centre for Cosmology, Particle Physics and Phenomenology, Institute of Mathematics and Physics, Louvain University, 2 Chemin du Cyclotron, 1348 Louvain-la-Neuve, Belgium \\
$^{(d)} $Namur Center of Complex Systems (naXys), Department of Mathematics, University of Namur, Rempart de la Vierge 8, 5000 Namur, Belgium \\
$^{(e)}$D\'epartment de Physique Th\'eorique, Universit\'e de Gen\`eve, 24 quai Ernest-Ansermet, 1211 Gen\`eve 4, Switzerland}

\preprint{CERN-TH-2017-271, IFT-UAM/CSIC-17-123}

\date{\today}

\begin{abstract}
It has been claimed in Ref.~\href{https://arxiv.org/abs/1712.02240}{\tt [arXiv:1712.02240]}, that massive primordial black holes (PBH) cannot constitute all of the dark matter (DM), because their gravitational-lensing imprint on the Hubble diagram of type Ia supernovae (SN) would be incompatible with present observations. In this note, we critically review those constraints and find several caveats on the analysis. First of all, the constraints on the fraction $\alpha$ of PBH in matter seem to be driven by a very restrictive choice of priors on the cosmological parameters. In particular, the degeneracy between $\OM$ and $\alpha$ was ignored and thus, by fixing $\OM$, transferred the constraining power of SN magnitudes to $\alpha$. Furthermore, by considering more realistic physical sizes for the type-Ia supernovae, we find an effect on the SN lensing magnification distribution that leads to significantly looser constraints.  Moreover, considering a wide mass spectrum of PBH, such as a lognormal distribution, further softens the constraints from SN lensing.  Finally, we find that the fraction of PBH that could constitute DM today is bounded by $\fpbh < 1.09\ (1.38)$, for JLA (Union 2.1) catalogs, and thus it is perfectly compatible with an all-PBH dark matter scenario in the LIGO band. 
\end{abstract}

\maketitle

\section{Introduction}

The detection of gravitational waves from the merging of massive black hole binaries by the Advanced LIGO/VIRGO interferometers~\cite{Abbott:2016blz,Abbott:2016nmj,TheLIGOScientific:2016pea,Abbott:2017vtc,Abbott:2017oio,Abbott:2017gyy} has revived the idea that primordial black holes (PBH) could constitute all (or an important fraction) of the dark matter.   The merging rates inferred by LIGO are indeed compatible with a black hole population having abundances similar to the one of dark matter~\cite{Bird:2016dcv,Clesse:2016vqa} (nevertheless see~\cite{Sasaki:2016jop} for a different conclusion), whereas their relatively large mass and low spins were mostly unexpected and might point towards a primordial origin.   These properties of the LIGO black hole binaries would come naturally from early universe models of PBH formation from large peaks in the matter power spectrum, arising both in single (e.g.~\cite{Garcia-Bellido:2017mdw,Ezquiaga:2017fvi}) and multi-field (e.g.~\cite{Garcia-Bellido:2016dkw,Garcia-Bellido:2017aan}) models of inflation, which leads to a wide mass distribution as well as to significant PBH clustering.

Motivated by the LIGO detections, the constraints on PBH abundances from the EROS microlensing survey~\cite{Tisserand:2006zx} and from the cosmic microwave background temperature anisotropies and spectral distortions~\cite{Ricotti:2007au}, have been reinvestigated.  On the one hand, EROS limits can easily be evaded~\cite{Hawkins:2011qz,Green:2017qoa,Garcia-Bellido:2017xvr}, e.g. when considering more realistic dark matter distributions in the galaxy or if most PBH are regrouped in micro-clusters, which reopened the low-mass ($M_{\rm PBH} \lesssim M_\odot$) window for PBH-DM.  Furthermore, other surveys of M31 stars and distant quasars found numerous microlensing events~\cite{Wyrzykowski:2015ppa,Mediavilla:2017bok}, which seems to contradict the EROS results.   On the other hand, more detailed computations of the  impact of early matter accretion on the CMB shows that Planck limits on PBH abundances are very sensitive to the PBH velocity with respect to baryons,  whereas there is no relevant constraint from CMB spectral distortions~\cite{Poulin:2017bwe,Ali-Haimoud:2016mbv}.   At the same time, new constraints have been established in this range $[10-100] M_\odot$ , especially from the dynamical heating of faint dwarf galaxies and their star clusters~\cite{Brandt:2016aco,Green:2016xgy,Li:2016utv}.    Altogether, the present status is that a wide PBH mass distribution centered on a few solar masses can perfectly pass all those limits and at the same time explain the LIGO events~\cite{Clesse:2017bsw}.  

The scenario of massive micro-clustered PBH~\cite{Garcia-Bellido:2017fdg} is supported by the non-detection of ultra-faint dwarf galaxies smaller than the critical dynamical heating radius~\cite{Clesse:2017bsw}, as well as by spatial correlations between the cosmic infrared and soft X-ray backgrounds~\cite{Kashlinsky:2016sdv,Cappelluti:2017nww}. It could explain the small-scale crisis linked to the core-cups, too-big-to-fail, missing satellites and missing baryons problems, thanks to PBH gravitational scattering and to the rapid gas accretion in dwarf galaxies~\cite{Clesse:2017bsw}.  Heavy PBH in the tail of the distribution would also provide the seeds for supermassive black holes~\cite{Clesse:2015wea} whose existence at high redshifts is still puzzling.

The recent claim by Zumalac\'arregui and Seljak in Ref.~\cite{Zumalacarregui:2017qqd} (hereafter [ZS17]) that the absence of lensing in type-Ia supernovae (SN) observations rules out PBH in the LIGO range $[0.01-100] \, \Msun$ as the main source of dark matter, even for wide mass distributions, would therefore strongly jeopardize the best motivated PBH-DM scenarios.  In this note, we carefully review the SN-Ia lensing constraints on PBH abundance and find several caveats on the analysis. First of all, the constraints on the fraction $\alpha$ of PBH in matter seem to be driven by a very restrictive choice of priors on the cosmological parameters. In particular, the well known degeneracy between $\OM$ and $\alpha$ is ignored and thus, by fixing $\OM$, they transfer the constraining power of SN magnitudes to $\alpha$. Furthermore, we point out that the characteristic size of SN is larger than the one assumed in [ZS17], which impacts the SN magnification PDF and softens the constraints by about one order of magnitude.   PBH can therefore have DM abundances, even in the simplest monochromatic case.  Furthermore, when considering a more realistic broad lognormal mass distribution, motivated by formation models from broad peaks in the power spectrum, we show that SN constraints on PBH abundances are further diluted.   Finally, we evaluate the effect of PBH clustering.  In this case, SN-Ia constraints are again more easily evaded.  We therefore conclude that few solar mass PBH-DM models are still consistent with observations. Nevertheless, we believe that the idea of using SN lensing to {\it detect} such PBH population is very promising and worth pursuing with future surveys like DES, LSST and Euclid.

In Section~\ref{sec:SNlensing}, we briefly introduce the SN lensing model and the caveats in the precedent approach.  Then, in Section~\ref{sec:lognormal}, we compute the SN constraints for a log-normal mass distribution of PBH.  Finally, we discuss our results and perspectives in Section~\ref{sec:ccl}.

\section{Supernova lensing}\label{sec:SNlensing}

\subsection{Model} \label{sec:SNlensingmodel}

SNe are gravitationally lensed by the matter density that their light encounter from emission to observation. This generates residuals in the observed Hubble diagram: some SNe appear brighter or fainter---i.e. closer or farther---than the average, assumed to be well-described by the FLRW model. This effect can be quantified by the magnification\footnote{Note that this convention for the magnification differs from the most common one~\cite{gravitational_lenses}, $\mu\e{st}=(D\e{FLRW}/D)^2$, defined with respect to an FLRW background (filled beam), and such that $\mu=1$ (instead of $0$) when there is no lensing.}
\begin{equation}
\mu \equiv \left(\frac{D\e{EB}}{D}\right)^2 - 1,
\end{equation}
where we adopted the convention of [ZS17], $D$ being the observed angular diameter distance to the SN, and $D\e{EB}$ the distance that would be observed if light would encounter no matter (``empty beam''~\cite{1964SvA.....8...13Z}).

Following Ref~\cite{Seljak:1999tm}, [ZS17] distinguishes between two effects: the weak-lensing magnification~$\mu\e{s}$ due to smooth matter encountered by the light beam, and the microlensing magnification $\mu\e{c}$ due to compact objects. If $\mu\e{s}\ll 1$, then the total magnification~$\mu$ reads
\begin{equation}
\mu = \mu\e{s} + \mu\e{c}
\end{equation}
If now $\mu'$ denotes the weak-lensing convergence \emph{in the absence of any compact object}, while a fraction~$\alpha$ of matter is actually made of such objects, then
\begin{equation}\label{eq:total_magnification}
\mu = (1-\alpha)\mu' + \mu\e{c}[\alpha\mu']
\end{equation}
where the dependence of $\mu\e{c}$ in $\alpha\mu'$ comes from the fact that the latter indicates the amount of compact objects along the line of sight; the larger $\alpha\mu'$ is, the more probable it is that light encounters such objects.

\begin{figure}[!ht]
\centering
\hspace*{-2mm}
\includegraphics[width = 0.49\textwidth]{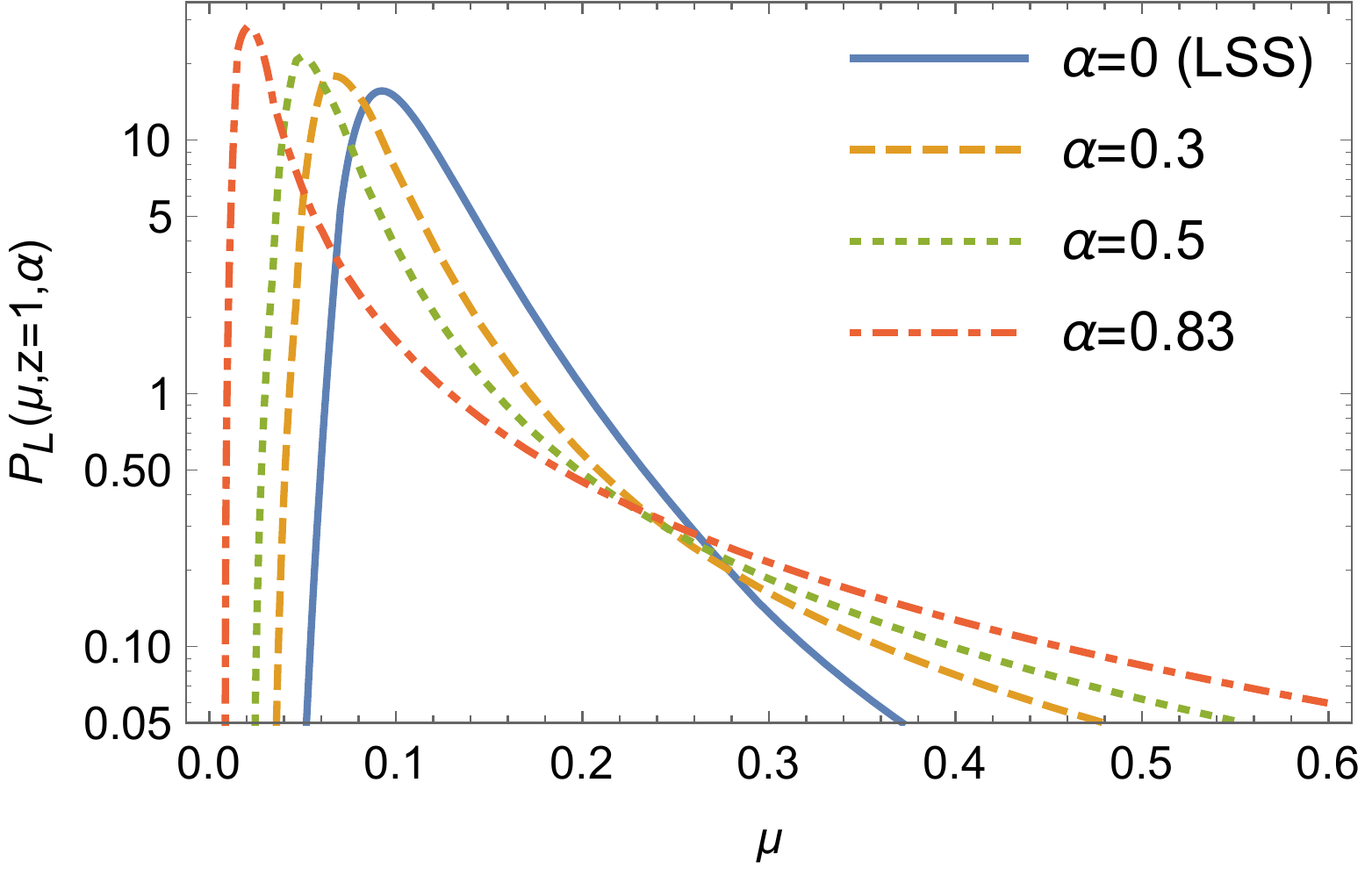}
\caption{PDF of the magnification, at $z=1$, for different values of the fraction~$\alpha$ of dark matter made of compact objects.}
\vspace*{-1mm}
\label{fig:PLz10}
\end{figure}

Equation~\eqref{eq:total_magnification} implies that the probability distribution function (PDF) of $\mu$ is a convolution the PDF of $\mu'$, due to the large-scale structure, and the PDF of $\mu\e{c}$, as
\begin{equation}\label{eq:convolution_PDFs}
P\e{L}(\mu; z,\alpha) = \int_0^{\frac{\mu}{1-\alpha}} \rr{d}\mu' \; 
P\e{LSS}(\mu';z) P\e{C}[\mu - (1-\alpha)\mu';\alpha\mu'].
\end{equation}
In [ZS17], $P\e{LSS}$ is generated by the code \texttt{turboGL}~\cite{Kainulainen:2009dw,Kainulainen:2010at} for a {\it Planck} cosmology and a halo mass function in the range $M\in[10^4,\,10^{16}]\,\Msun$. Besides, $P\e{C}$ is a fitting formula to numerical simulations performed by Rauch~\cite{1991ApJ...383..466R}, in a universe filled with a uniform comoving density of compact objects, with no cosmological constant,
\begin{equation}\label{eq:fitting_formula_PC}
P\e{C}(\mu,\bar{\mu}) = A\,\left[\frac{1-\rr{e}^{-\mu/\delta}}{(\mu+1)^2-1}\right]^{3/2}
\quad \text{for }\mu >0\,,
\end{equation}
and 0 otherwise. The parameters $A$ and $\delta$ are determined by enforcing that $P\e{C}$ is normalized, and that $\langle \mu \rangle = \bar{\mu}$. This latter condition corresponds to the assumption that lensing by small-scale structures does not affect the average magnification~\cite{Weinberg:1976jq}, which remains debated in the lensing community~\cite{Fleury:2015hgz}. The result is plotted in Fig.~\ref{fig:PLz10} for $\alpha=0,0.3,0.5,0.83$.

Last, but not least, [ZS17] took into account that SNe are not exactly point-like sources. More precisely, their (unlensed) angular size~$\theta\e{S}$ is not always much smaller than the Einstein radius~$\theta\e{E}$ of the compact lenses, namely
\begin{equation}\label{eq:ratio_angular_sizes}
\frac{\theta\e{S}}{\theta\e{E}} = 1.11\,\frac{R\e{SN}}{100\,{\rm AU}}\,
\left(\frac{\Msun}{M}\,\frac{D\e{L}\,{\rm Mpc}}{D\e{S}\,D\e{LS}}\right)^{1/2}\,,
\end{equation}
where $R_S\sim100$ AU is the physical size of a SN during the observed phase of the event, and distances $D$ are angular diameter distances in units of Mpc. When $\theta\e{S}\ll \theta\e{E}$ (infinitesimal beam) the compact object can act as a strong gravitational lens, and hence enters into the $\mu\e{c}$ category; otherwise the light beam associated with the SN has to be considered finite~\cite{Fleury:2017owg}, and the lens effectively counts as a $\mu\e{s}$.

As discussed in [ZS17], this leads to a renormalization of the fraction of matter which must be counted as compact objects. If $\alpha \equiv \rho\e{PBH}/\rho\e{M}=0.83 f\e{PBH}$, where $f\e{PBH}\equiv\rho\e{PBH}/\rho\e{DM}$ denotes the fraction of dark matter in the form of PBH, then the fraction of associated strong lenses reads
\begin{equation}
\tilde{\alpha} = f\e{L} \alpha = 0.833 f\e{L} f\e{PBH}.
\end{equation}
This fraction depends on the redshift of the source~$z\e{S}$ and on the PBH mass~$M$ as
\begin{equation}\label{eq:fLM}
f\e{L}(z\e{S}, M)
=  \frac{\int_0^{z\e{S}}\rpbh(z)\,\Theta(\epsilon-\theta\e{S}/\theta\e{E})\;\rr{d}z}
		{\int_0^{z\e{S}}\rpbh(z)\,\rr{d}z}
\end{equation}
where, following [ZS17], we considered that a PBH is considered a strong lens if~$\theta\e{S}/\theta\e{E}<\epsilon=0.05$, $\Theta$ being the Heaviside distribution. Using that $\theta\e{S}/\theta\e{E}$ grows with the redshift of the lens, we can rewrite Eq.~\eqref{eq:fLM} as
\begin{equation}
f\e{L}(z\e{S}, M) = \frac{(1+z_*)^4-1}{(1+z\e{S})^4-1},
\end{equation}
with $z_*$ determined by $\theta\e{S}/\theta\e{E}=\epsilon$, and assuming that most of the PBH do not accrete mass n the last few e-folds of cosmic expansion $(z<2)$, and hence dilute with the volume, $\rpbh(z)\sim(1+z)^3$.

\subsection{Caveats}\label{sec:caveats}

The model presented in Sec.~\ref{sec:SNlensingmodel} corresponds to the state of the art for SN lensing by compact objects. [ZS17] applied it to two SN data sets, namely Union 2.1~\cite{Suzuki:2011hu}, and the Joint Lightcurve Analysis (JLA)~\cite{Betoule:2014frx}. They found that the distribution of SN magnitudes excludes that PBH with masses above~$10^{2}M_\odot$ constitute more than $30\%$ of the dark matter. There are, however, two important caveats in this analysis, on which we elaborate below. 

\subsubsection{Effective demagnification and $\Omega\e{m}$}

\begin{figure}[t]
\centering
\hspace*{-2mm}
\includegraphics[width = 0.49\textwidth]{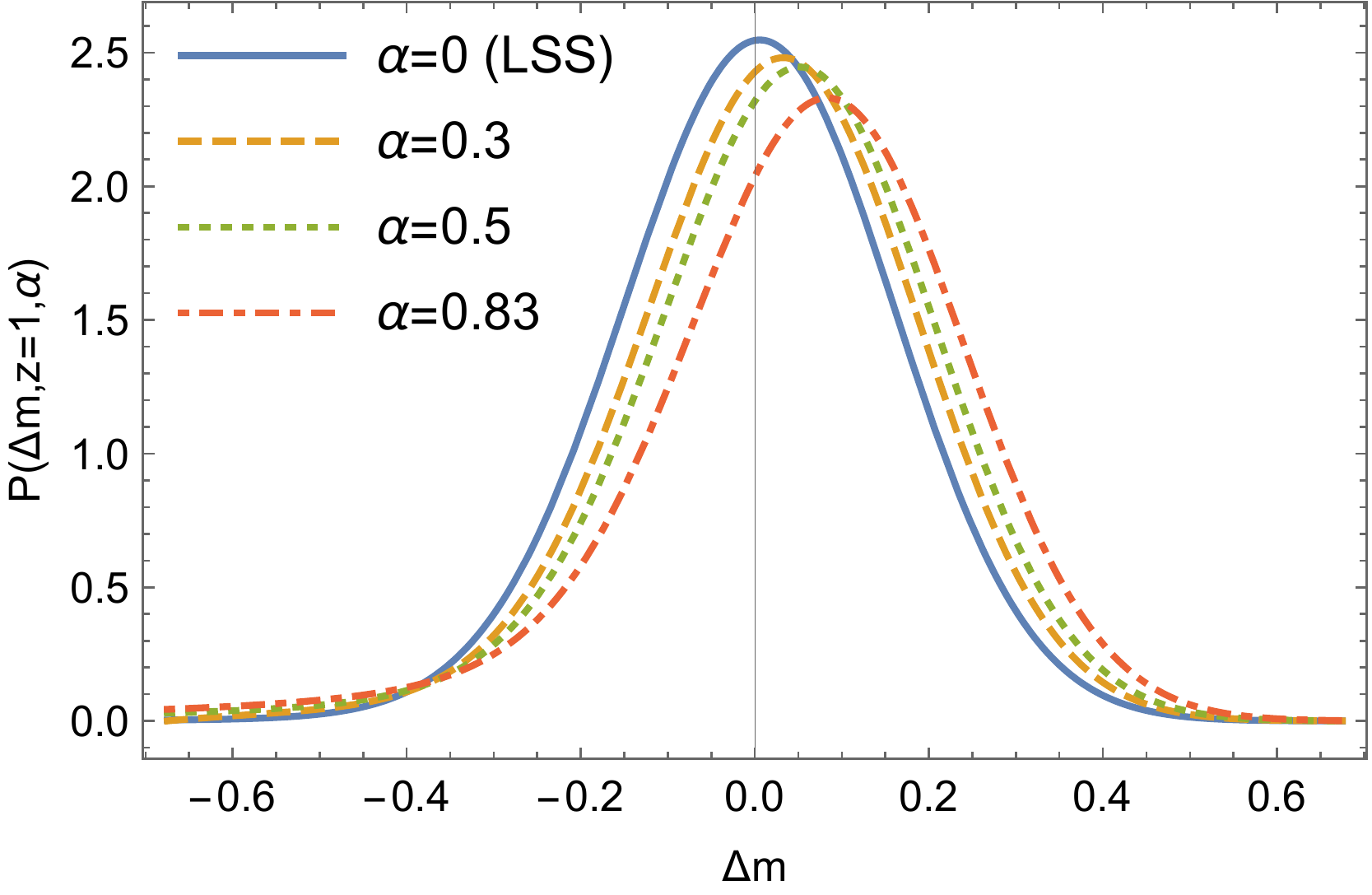}\\[3mm]
\includegraphics[width = 0.49\textwidth]{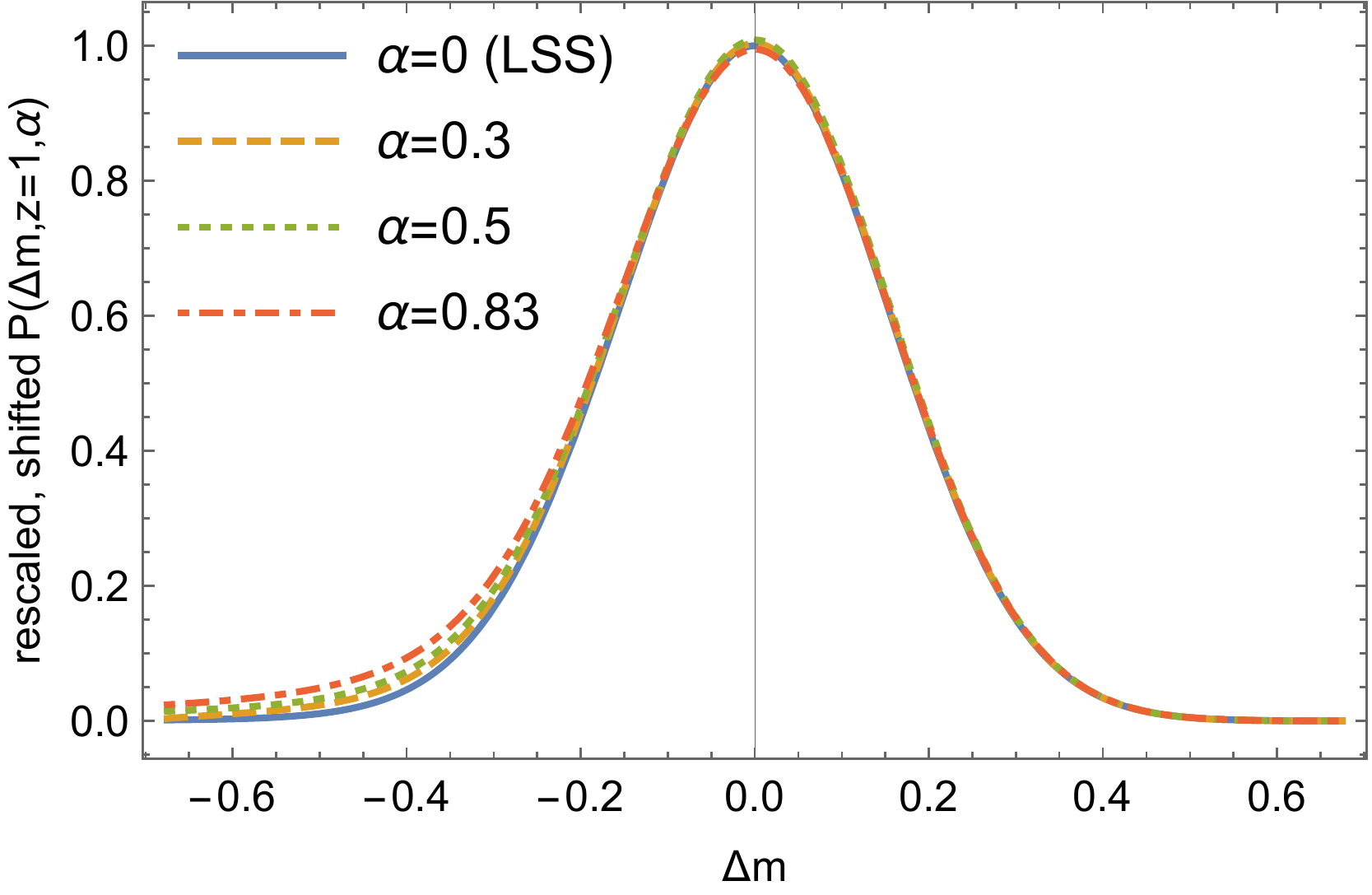}
\caption{Top: PDF of the magnitude of lensed SNe at $z=1$, including the intrisinc disperson of luminosities, for different values of the fraction~$\alpha$ of compact objects constituting the dark matter. Bottom: same as above but rescaled and shifted so that the maxima coincide.}
\label{fig:magnitudes}
\end{figure}

The net effect of lensing on the distribution of SN magnitudes can be observed in the top panel of Fig.~\ref{fig:magnitudes}, where $\Delta m$ denotes the magnitude shift with respect to the mean. To produce this distribution, we allowed for the fact that, irrespective of lensing, SNe have an intrinsic scatter in their luminosity due to various effects, from environment issues like metallicity to misidentification of hosts. This scatter is commonly modelled by a Gaussian distribution in magnitude, with $\sigma_m = 0.15$, assumed to be redshift independent~\cite{Betoule:2014frx}. Besides, the magnitude shift due to a magnification~$\mu$ reads
\begin{equation}
\Delta m = -\frac{5}{2} \log_{10}(1+\mu - \mu\e{F}),
\end{equation}
where $\mu\e{F}\approx 0.12$ is the magnification of the ``filled beam'' (FLRW) with respect to the ``empty beam''. The predicted PDF of $\Delta m$ is thus obtained by convolving the intrinsic distribution with $P\e{L}(\Delta m)=|\dd\mu/\dd\Delta m| P\e{L}(\mu)$.

As seen in Fig.~\ref{fig:magnitudes}, the most significant effect of lensing is to move the most probable magnitude towards larger values. The predominance of this shift can be made even clearer by translating and rescaling all the distributions, such that the maxima coincide (bottom panel of Fig.~\ref{fig:magnitudes}). The resulting curves are almost identical, and it is hard to believe that the current SN data is able distinguish between $\alpha=0$ and $\alpha=0.83$ from their shapes only.

On the contrary, SNe are quite efficient at measuring the behaviour of the mean magnitude-redshift relation. Of course, in theory, all the distributions for the top panel of Fig.~\ref{fig:magnitudes} have, by construction, the same mean---the tail for negative values of $\Delta m$ exactly compensates for the shift of the peak. Nevertheless, in practice, this reasoning only applies if there is enough data to sample the whole distribution, including the tails. This is not the case for current SN surveys, characterized by a relatively low number of objects (580 for Union 2.1 and 740 for JLA). Therefore, the dominant effect of increasing $\alpha$ is to bias up the average magnitude-redshift relation.

This effect is essentially equivalent to reducing the mean density of the Universe~\cite{Clarkson:2011br,Fleury:2013sna,Fleury:2013uqa} (or adding spatial curvature). Indeed, the denser the Universe, the more matter is encountered by light beams, and hence the more magnified they are. Hence, in the analysis of SN data, there should be an important degeneracy between $\Omega\e{m}$ and $\alpha$. This issue was overlooked in [ZS17], where a prior on $\Omega\e{m}$ ($0.309\pm 0.006$) was set from CMB+BAO data; since this prior is much narrower than the constraints on $\Omega\e{m}$ set by SN data alone, this artificially transfers the constraining power of SN data to $\alpha$. In other words, it is not excluded that the constraints on $\alpha$ obtained by [ZS17] are driven by this prior.

This hypothesis is supported by the absence of skewness signal in the constraints of [ZS17]. Indeed, the larger $\alpha$, the more skewed the distribution of magnitudes (see Fig.~\ref{fig:magnitudes}). Thus, one could expect this skewness to the dominant lensing signal to look for in the SN data. If it were the case, however, the lensing skewness should be hardly distinguishable from the skewness of the distribution of intrinsic SN luminosities, quantified by the parameter~$k_3$ in the analysis of [ZS17]. In other words, there should be a degeneracy between $k_3$ and $\alpha$ in the constraints set by the data. Yet this is not observed in Fig.~6 of [ZS17].

\subsubsection{Misestimation of the fraction of strong lenses}

Even if one neglects the above issue, and accept the methodology of [ZS17], the resulting constraints still turn out to be weaker than what was claimed, because [ZS17] underestimated the ratio $\theta\e{S}/\theta\e{E}$ of Eq.~\eqref{eq:ratio_angular_sizes} by a factor $50$. As a consequence, the condition $\theta\e{S}/\theta\e{E}<\epsilon$ is more restrictive in reality than it is in [ZS17], so less PBH contribute as compact lenses, and hence a given constraint on $\alpha$ translates into a weaker constraint on $f\e{PBH}$. In particular, we will show in the next section that, for PBH with a lognormal distribution of mass, SN lensing ends up not constraining $f\e{PBH}$ at all.

One could circumvent this issue by arguing that $\epsilon = 5\%$ is a very conservative bound, as already suggested in [ZS17]. In fact, it is hard to guess what a sensible threshold is for the transition between strong and weak lensing; but the sensitivity of the constraints with respect to the estimate of $\theta\e{S}/\theta\e{E}$ (and hence the choice of $\epsilon$) shows that this point must be taken seriously. This calls for more elaborate methods to estimate the effective fraction of compact lenses in a given model for an arbitrary matter distribution.

\section{Lognormal mass spectrum}  \label{sec:lognormal}

In this section, we show that, even if we neglect the first point of Sec.~\ref{sec:caveats}, reproducing the analysis of [ZS17] with a corrected value for $\theta\e{S}/\theta\e{E}$ leads to constraints in full agreement with PBH constituting all of the dark matter, when taking into account a realistic broad distribution of masses.

\subsection{Finite size clusters of PBH}

We still do not know how PBHs are distributed in space. Depending on their production mechanism, they could have arisen from broad peaks in the matter power spectrum that collapsed to form black holes in the radiation era. In this case, PBH come in clusters of variable numbers, from 100 to 1000 PBH in each cluster, and subtending a substantial transverse dimension, of order a milliparsec. We ignore how these clusters have evolved since recombination. It could be that most of them have ``evaporated", heated up by dynamical interactions among the members of the cluster~\cite{Sigurdsson:1993zrm}, or merged in the presence of gas to form the more massive IMBH and SMBH at the centers of galaxies~\cite{Clesse:2015wea}.

\begin{figure}[!ht]
\centering
\includegraphics[width = 0.48\textwidth]{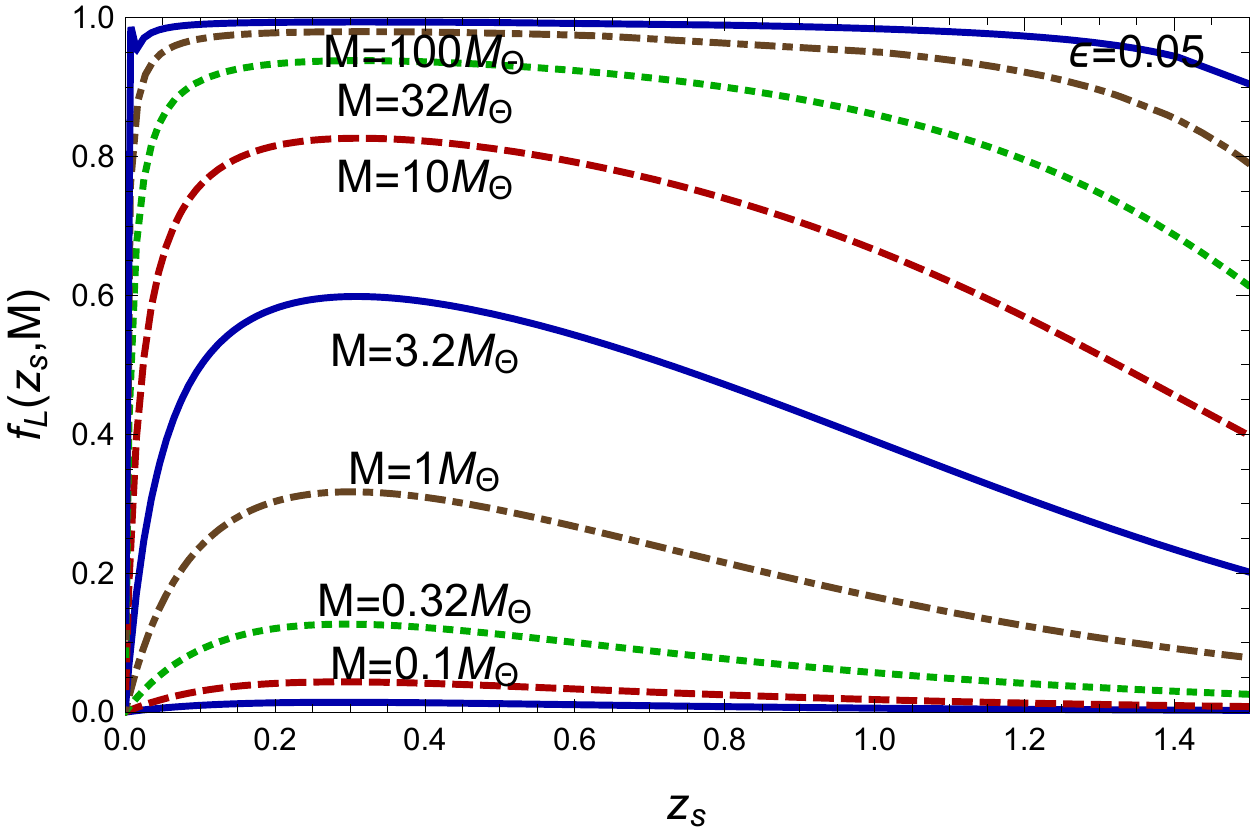}\\
\includegraphics[width = 0.48\textwidth]{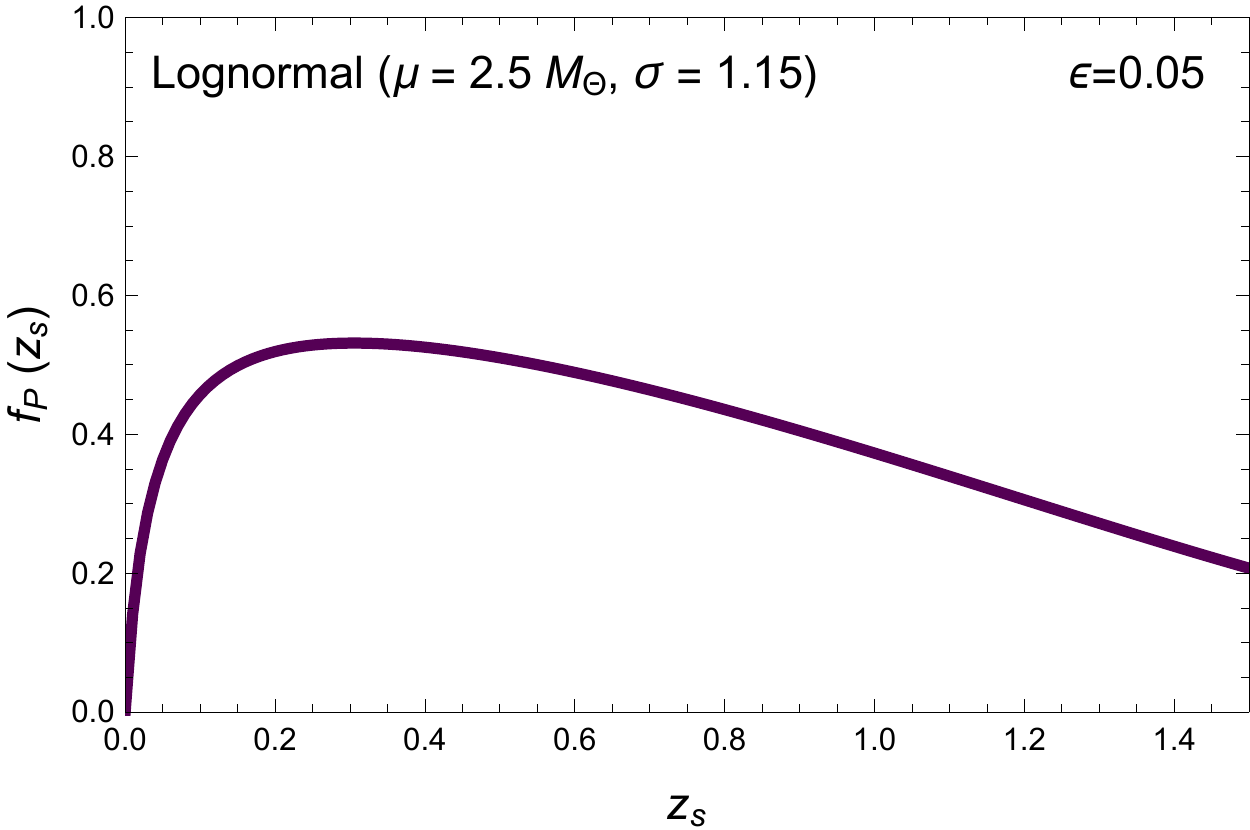}
\caption{Top: The effective PBH fraction as a function of the source redshift for different PBH masses. Bottom: The mass-integrated PBH fraction as a function of the source redshift, for a Lognormal distribution with parameters $\mu=2.5\,\Msun$ and $\sigma=1.15$, as found in agreement with all present constraints~\cite{Clesse:2017bsw}.}
\vspace*{-1mm}
\label{fig:effective}
\end{figure}

Therefore, it is relevant to consider the fraction of them that permeate the universe and constitute the CDM that helped baryons form structures like galaxies and clusters of galaxies. These objects have masses well below the resolution of N-body simulations and therefore can be considered as the smooth component of CDM, forming the DM halos and moving in the potential wells of long-wavelength curvature fluctuations set up by inflation. However, if the PBH hypothesis is correct, these halos are composed of compact objects and these may act as lenses. It remains to be studied the effect that these hard cores have on the total (weak, strong and micro) lensing of distant supernovae.

\subsection{PBH Lognormal distribution}

So far we have assumed that PBH have a single mass. However, even if they had started having a monochromatic mass spectrum, accretion would inevitably have spread their mass spectrum. Alternatively, they could have arisen from broad peaks in the matter power spectrum and thus have a wide distribution to start with. Whatever the origin, we can assume, for simplicity, that the PBH mass distribution is lognormal with parameters ($\mu,\,\sigma$),
\be\label{eq:LN}
P(M) =  \frac{\fpbh}{M\,\sqrt{2\pi\,\sigma^2}}\,\exp\left[-\frac{\ln^2(M/\mu)}{2\sigma^2}\right]\,,
\ee
where we have chosen $P(M)$ to be normalized to the $\fpbh$ fraction, $\int_0^\infty P(M) \,\rr{d}M = \fpbh$. At present, there are several hints that suggest that PBH constitute all the DM~\cite{Clesse:2017bsw}, from the observed LIGO events to the formation of structure in the early universe, and the best fit values for the parameters of Eq.~(\ref{eq:LN}) are
\be
\mu = 2.5\,\Msun\,, \qquad \sigma = 0.5\,\ln10\,,
\ee
which is a relatively wide distribution. In this case, one can compute the effective PBH fraction that may act as lens, as a function of the redshift
\be\label{eq:azs}
\tilde{\alpha}(z\e{S}) = \alpha f\e{P}(z\e{S}) = 
\alpha \int_0^\infty P(M) f\e{L}(z\e{S}, M) \;\rr{d}M\,.
\ee
We have plotted in Fig.~\ref{fig:effective} (bottom panel) the effective fraction $f\e{P}(z\e{S})$ after integration of the lognormal distribution with the above parameters. It is clear that the fraction becomes significantly smaller than one for a wide range of redshifts.

\begin{figure}[!t]
\centering
\includegraphics[width = 0.48\textwidth]{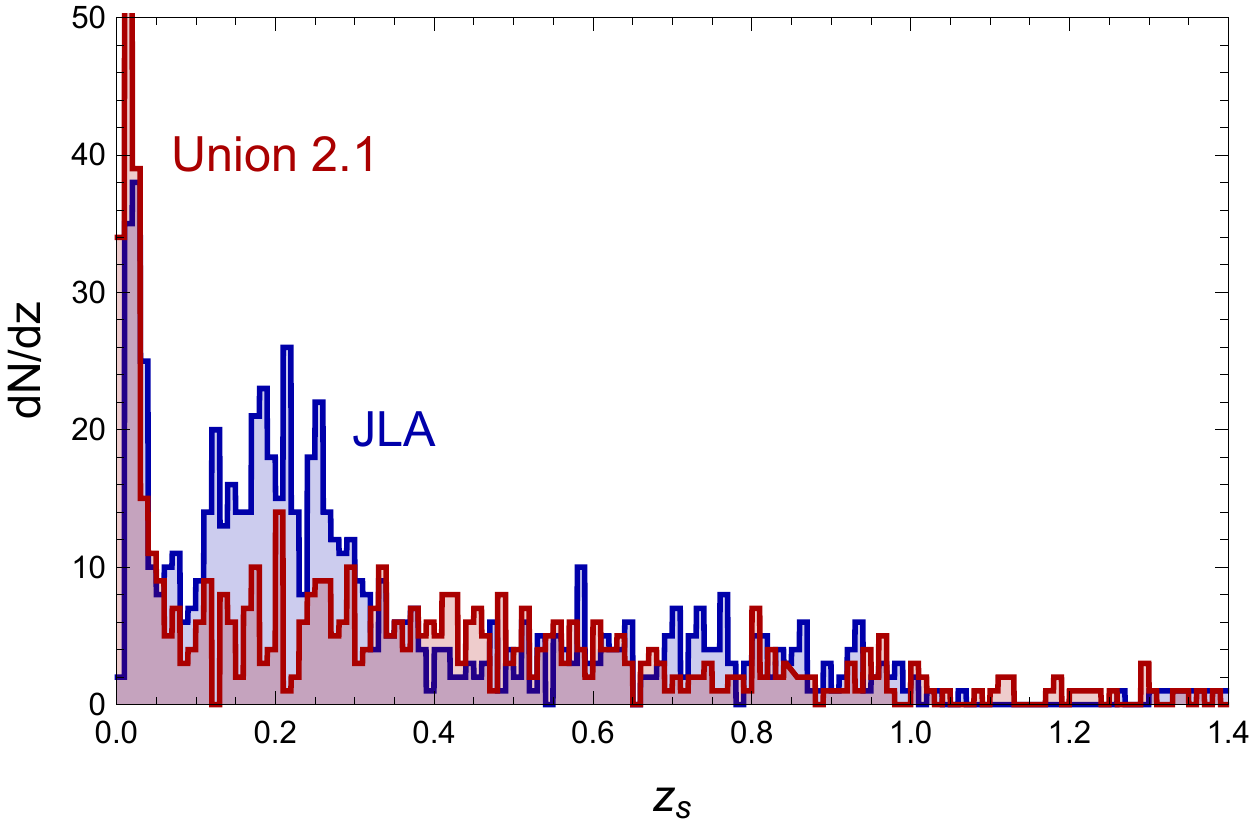}
\caption{The JLA and Union 2.1 SN-Ia redshift distributions.}
\vspace*{-1mm}
\label{fig:JLA}
\end{figure}

\subsection{Supernova redshift distribution}

As $\tilde{\alpha}$ depends on redshift, we must integrate it over the redshift distribution of the SNe, in order to derive its effective value for the survey at hand. We will assume that the lensing effect of PBH on SN-Ia magnitudes is independent of redshift (see the previous footnote for a caveat), and that each redshift bin contributes equally to the final effective value of the PBH fraction,
\be
\tilde{\alpha}_{\rm eff} = \alpha\ \int f\e{P}(z)\,\frac{\rr{d}N}{\rr{d}z}(z)\;\rr{d}z\,.
\ee
We consider both JLA and the Union 2.1, whose redshift distributions~$\rr{d}N/\rr{d}z$ are depicted in Fig.~\ref{fig:JLA}, and we obtain $\alpha_{\rm eff}^{\rm JLA} = 0.44\,\alpha$ for the JLA catalog and $\alpha_{\rm eff}^{\rm Union} = 0.39\,\alpha$ for the Union 2.1 catalog. This means that the overall constraint on the fraction of PBH in DM is given by
\be
\fpbh = \alpha_{\rm eff} \times 
\begin{cases}
2.73 &\quad {\rm JLA}\\[2mm]
3.08 &\quad {\rm Union 2.1}
\end{cases}
\ee

Using the $95\%$ bounds $\alpha < 0.4$ (JLA), and $\alpha < 0.45$ (Union2.1) obtained by [ZS17], which allow for the full non-diagonal covariance matrix of the surveys, the constraints on $\fpbh$ become
\be\label{eq:cons}
\fpbh < 
\begin{cases}
1.09 &\quad {\rm JLA}\\[2mm]
1.38 &\quad {\rm Union 2.1}
\end{cases}
\ee
Therefore, one hundred percent of DM in the form of PBH is fully consistent with SN lensing.

\begin{figure}[!ht]
\centering
\includegraphics[width = 0.48\textwidth]{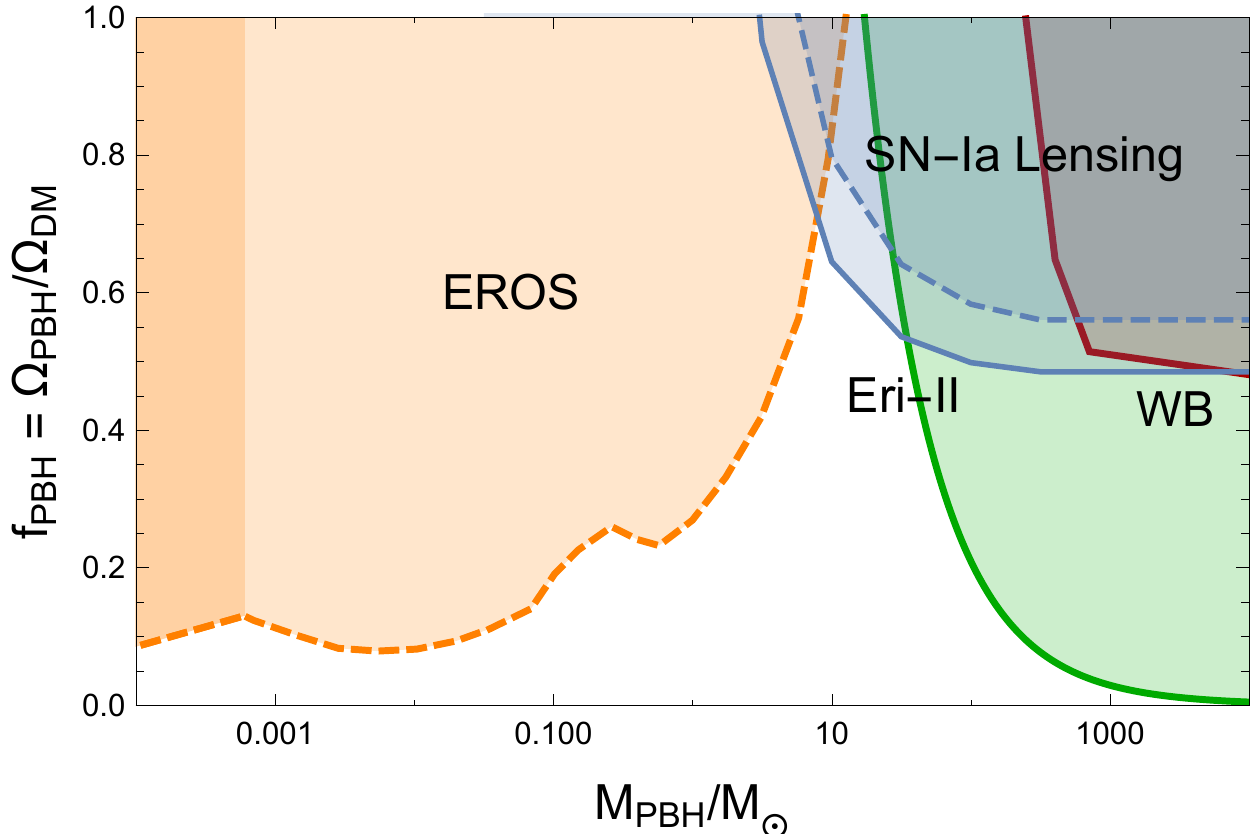}
\caption{The constraints on PBH-DM in the range of masses near the LIGO band. We have shown with a dashed line the constraints from EROS because of the assumption of monochromaticity and the absence of clustering, which essentially removes this constraint from the band, see Ref.~\cite{Garcia-Bellido:2017xvr}. The blue contours correspond to the SN lensing constraints for JLA (continuous) and Union 2.1 (dashed) SN catalogs; the green contours arise from Eridanus II dwarf spheroidal~\cite{Brandt:2016aco,Green:2016xgy,Li:2016utv}, and the red contours from the disruptiuon of wide binaries in the Milky Way halo~\cite{MonroyRodriguez:2014ula}.}
\vspace*{-1mm}
\label{fig:PBHcons}
\end{figure}

\section{Conclusions}  \label{sec:ccl}

We have reanalyzed the claims of [ZS17] for ruling out PBH-DM with masses in the LIGO range due to the lack of lensing of type-Ia supernovae.  First, it has been put in evidence that the relatively small number of SNe induces an effective bias in the mean magnitude that is degenerated with $\Omega_{\rm m}$.   As a result, the constraint on the PBH fraction could be partially explained by the choice of the prior limits for $\Omega_{\rm m}$.  Second, the characteristic size of SNe was underestimated in  [ZS17], which impacts the SN lensing magnification distribution and softens the constraint on the PBH abundance.   Even in the case of a monochromatic spectrum, SNe allow PBH-DM for masses lower than about $3 M_\odot$, as illustrated on Fig.~\ref{fig:PBHcons}, and opposite to the claim of  [ZS17].   The SN-Ia lensing constraints on Fig.~\ref{fig:PBHcons} were obtained by integrating Eq.~(\ref{eq:azs}) over the redshift distribution of JLA (Union 2.1)
\be\label{eq:aM}
\alpha(M) = \alpha\ \int f\e{L}(z,M)\,\frac{\rr{d}N}{\rr{d}z}(z)\;\rr{d}z\,,
\ee
and then taking the constraints on $\alpha$ from [ZS17] and inverting to get the constraints on $\fpbh$.   This limit improves the constraint from Eridanus II by a factor four (more than ten if one assumes an intermediate-mass black hole at the center of ERI-II).  Given that microlensing constraints from EROS are uncertain and in tension with other surveys, the monochromatic PBH-DM scenario in the sub-solar range still survives all the constraints.   

In a more realistic scenario, one needs a relatively wide mass distribution to explain the LIGO merger events, e.g. a log-normal distribution centered in the range $[1-10]\, M_\odot$~\cite{Garcia-Bellido:2017fdg}. In this case, we have shown that the SNe lead to further loser constraints, see Eq.~(\ref{eq:cons}), that do not exclude PBH to be all of the dark matter.  Finally,  micro-clustering of PBH, like in the scenario of~\cite{Garcia-Bellido:2017fdg}, would also lead to loser constraints.  Note that the EROS constraints assume a monochromatic distribution, and that integrating over a lognormal distribution makes them wider but less restrictive, see Ref.~\cite{Garcia-Bellido:2017xvr}, contrary to what is claimed by [ZS17]. Nevertheless, what is relevant is the global constraint on the fraction of PBH in DM, see Eq.~(\ref{eq:cons}), which indicates that dark matter could be entirely comprised of PBH of masses within the LIGO band, contradicting the strong claims made in [ZS17].

Note that a similar methodological approach to the problem of SN lensing by small scale structure was taken in Refs.~\cite{Bergstrom:1999xh,Mortsell:2001}, and shown to help the overall agreement of the concordance $\Lambda$CDM paradigm in Refs.~\cite{Amendola:2010ub,Amendola:2013twa}. Moreover, that galaxies and smaller structures can actually lense type-Ia supernovae has recently been shown in Refs.~\cite{Quimby:2014irr,Goobar:2016uuf} with explicit examples. It will therefore not come as a surprise if in the future we will discover a whole population of lensed type-Ia SN. In fact, the next generation of supernova surveys (DES, LSST, Euclid) will have significantly more statistics and thus a better control of systematics, which will provide a real opportunity to search for compact objects like PBH as the dominant contribution of DM.

\section*{Acknowledgements}

We thank Masao Sako, Valerio Marra, Savvas Nesseris and Ariel Goobar for comments and discussions. JGB thanks the CERN TH-Division for hospitality during his sabbatical, and acknowledges support from the Research Project FPA2015-68048-03-3P [MINECO-FEDER] and the Centro de Excelencia Severo Ochoa Program SEV-2012-0597. He also acknowledges support from the Salvador de Madariaga Program, Ref. PRX17/00056. The work of SC is supported by a \textit{Charg\'e de Recherche} grant of the Belgian Fund for Research FRS/FNRS. PF acknowledges support by the Swiss National Science Foundation.

\bibliography{biblio}

\end{document}